# Research on the Design of a Short Video Recommendation System Based on Multimodal Information and Differential Privacy

Haowei Yang*

University of Houston, hyang38@cougarnet.uh.edu

Lei Fu

Independent Researcher, fuleiac@gmail.com

QINGYI LU

Brown University, lunalu9739@gmail.com

Yue Fan

Case Western Reserve University, yxf486@case.edu

Tianle Zhang

Independent Researche ,tianle.zhang@hotmail.com

Ruohan Wang

Johns Hopkins University,ruohanww@gmail.com

With the rapid development of short video platforms, recommendation systems have become key technologies for improving user experience and enhancing platform engagement. However, while short video recommendation systems leverage multimodal information (such as images, text, and audio) to improve recommendation effectiveness, they also face the severe challenge of user privacy leakage. This paper proposes a short video recommendation system based on multimodal information and differential privacy protection. First, deep learning models are used for feature extraction and fusion of multimodal data, effectively improving recommendation accuracy. Then, a differential privacy protection mechanism suitable for recommendation scenarios is designed to ensure user data privacy while maintaining system performance. Experimental results show that the proposed method outperforms existing mainstream approaches in terms of recommendation accuracy, multimodal fusion effectiveness, and privacy protection performance, providing important insights for the design of recommendation systems for short video platforms.

---

* Place the footnote text for the author (if applicable) here.

## 1 INTRODUCTION

With the rapid development of internet technology, short video platforms have gradually become an important channel for users to access information and entertainment. Short video content is characterized by efficient dissemination, personalized display, and strong interactivity, attracting a large number of users and creators. In this context, recommendation systems have played a crucial role in enhancing user experience, increasing platform engagement, and optimizing content distribution on short video platforms. However, existing short video recommendation systems still face many challenges in their technical implementation.First, short video content typically includes multimodal information such as images, text, and audio, and the effective fusion of these modalities significantly impacts the accuracy and diversity of recommendation systems. However, the heterogeneity and complexity of multimodal data make feature extraction and fusion a major technical challenge. Additionally, as privacy protection regulations become stricter and user privacy awareness increases, recommendation systems must balance privacy protection with recommendation effectiveness when processing user behavior data[1]. This poses higher requirements for the design of short video recommendation systems.To address these issues, this paper proposes a short video recommendation system based on multimodal information and differential privacy protection. By extracting and fusing deep features from multimodal information such as images, text, and audio, the accuracy and applicability of the recommendation algorithm are improved[2]. Meanwhile, the introduction of a differential privacy protection mechanism strengthens the system's ability to protect user data privacy while achieving an effective balance between recommendation effectiveness and privacy protection[3]. This research provides a new solution for recommendation technology on short video platforms and offers valuable insights into the application of privacy protection technologies in recommendation systems. In related work, Zixiang Wang et al. [4]studied deep reinforcement learning for autonomous driving decision-making. Haoqi et al.[5] proposed a texture-based method for sedimentary structure recognition. Qiming Yang et al.[6] improved U-Net for remote sensing image segmentation.Ao Xiang et al.[7] proposed a multimodal emotion recognition network to help me analyze students' emotions.Mingwei Wang et al.[8] studied teen online learning with machine learning. Yihao Zhong et al.[9] compared deep learning models for pneumonia detection. Changsong Wei et al. [10]explored explainable AI applications in natural language processing. Jiabei Liu et al.[11] proposed an out-of-distribution detection method. Gu Wenjun and his team [12] invented a new way to predict stock prices. Zong Ke [13-14] and his team also worked on detecting deep forgery and fraud in AI payments. They also explored using a backpropagation-based genetic algorithm to predict stock market volatility..

## 2 SYSTEM DESIGN AND ARCHITECTURE

### 2.1 Overall Architecture Design

The system is made up of several parts. These are a video feature extraction module, a user preference modelling module, a candidate video retrieval module, a candidate video ranking module and a recommended video output module. Together, these parts aim to efficiently provide personalised recommendations to users [15-17].

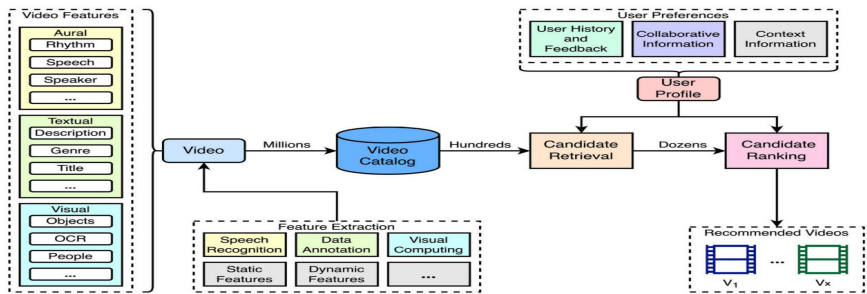

Figure 1: Overall Architecture of the Short Video Recommendation System

In the subsequent sections, we will elaborate on the specific methods for multimodal information fusion and the implementation of differential privacy protection strategies.Pei-Chiang Su, et al. [26]developed an algorithm for task scheduling. Ziqing Yin, et al.[27] compared CatBoost and XGBoost for turnover. Shaoxuan Sun, et al.[28] evaluated baseball teaching with ML. Chenming Duan, et al.[29] predicted athletes' mental states.

The video feature extraction module does three things. First, it takes in short videos and finds features in the audio, text and visual parts of the video. Then, it combines these features. Finally, it puts all of these features together to make a library of features. The audio includes things like rhythm, voice and speaker. The text includes things like the title, category and description of the video. And the visual includes object recognition, text that has been read out and facial recognition[18]. The feature extraction methods used include speech recognition, data tagging and visual computing. These methods generate static and dynamic features to support the recommendation phase [19]. The user preference modelling module uses data about past behaviour (e.g. clicks, likes, comments, etc.) and feedback, as well as information about the current situation, to create user profiles [20-22]. This module then uses this information to improve the accuracy and timeliness of recommendations. The video retrieval module looks for videos that might be interesting to the user in a large video catalogue. This module uses feature matching and fast indexing techniques to filter hundreds of candidate videos from millions of videos [23]. The next part, the candidate video ranking module, looks at the videos and combines information from different sources to decide which videos are likely to be the most popular. You can improve the effectiveness of recommendations by using deep learning models, such as Transformer-based ranking networks [24]. The recommended video output module recommends the sorted video content (e.g., V1 to Vx) to users to meet their different needs and optimise the content distribution of the platform. The most important part of this design is making the most of the features of multimodal information, while adding ways to protect the privacy of user data in each part of the system [25].In the next sections, we will explain how to combine different types of information and how to protect people's privacy. Jiangsu Pei and others developed a task scheduling algorithm.Yin Ziqing and others compared the turnover rates of CatBoost and XGBoost. Sun Shaoxuan et al [28] used Duan Chenming et al [29] to predict how athletes are feeling and evaluate baseball teaching.

## 2.2 Multimodal Information Fusion Module

The most important part of short video recommendation systems is accurately understanding what users like and need [30-32]. The multimodal information fusion module combines content features and user behavioural features to create a more comprehensive recommendation strategy [33]. Figure 2 shows different ways multimodal information is used in recommender systems. These include content similarity-based recommendations, collaborative user filtering, hybrid recommendation strategies and group preference fusion. These solutions are effective for different recommendation needs.

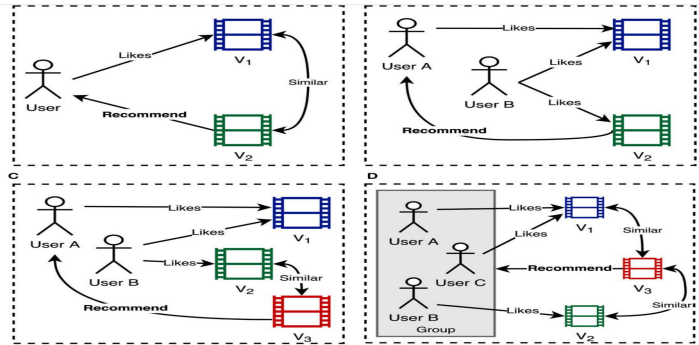

Figure 2: Illustration of the Application of Multimodal Information Fusion in the Recommendation System

*2.2.1 Content-based Recommendation*

As shown in Fig. 2(A), content-based recommendation focuses on the multimodal features of the video itself [34]. The system looks at the visual, textual, and audio parts of the video, such as images, written descriptions, and audio rhythms, to represent the video [35]. Then, based on the user's "like" behaviour of video V1, the system calculates the feature similarity and finds the video V2 that is most similar to V1 in the content features to achieve a personalised recommendation. For example, if video V1 is a dance performance, the system will recommend video V2 that has a similar rhythm or scene to attract users to watch related content [36].

*2.2.2 Collaborative Filtering-based Recommendation*

Figure 2(B) shows how collaborative filtering works. This is a system that recommends content based on how similar users are in terms of their behaviour. If user A is really into video V1 and user B has a similar taste, the system thinks that their interests are the same [37]. The system then looks at what else user B likes, like video V2, and recommends it to user A. This uses the knowledge of the crowd to find interesting points, which works well when there is a lot of data about user behaviour.

*2.2.3 Hybrid Recommendation Strategy*

As shown in Figure 2(C), the hybrid recommendation strategy combines content similarity and collaborative filtering to improve the comprehensiveness and diversity of recommendations [38]. In this case, if User A and User B both like Video V1 and User B likes Video V2, the system will recommend Video V2 to User A. In addition, the system may also recommend Video V3 if Video V2 has a strong similarity with Video V3 in terms of content features (e.g. visual objects or audio features).The strategy not only considers the collaborative relationships between users, but also exploits the connections between content features to generate a more hierarchical recommendation list.

*2.2.4 Group Preference Fusion Recommendation*

Figure 2(D) illustrates the application of group preference analysis in recommendation. When users A, B and C form a group, the system integrates the behavioural data of all users in the group and extracts common points of interest. For example, if user A likes video V1, user B likes video V2, and user C likes video V3, the system will find videos that satisfy the group's interests based on the content similarity of V1, V2, and V3, and make recommendations. This approach addresses the need for content recommendation in multi-user scenarios such as family or social groups.

The module uses deep learning techniques to extract features and unify multimodal information. For example, Convolutional Neural Networks (CNN) are used to extract visual features from videos, Natural Language Processing (NLP) models are used for textual modal features, and audio modal features are derived from spectral analysis techniques. A unified multimodal feature representation is constructed. In addition, in the feature fusion stage, the system uses an attention mechanism or a multimodal interaction network (MMIN) to weight and process different modalities to ensure that each modality contributes the most to the recommendation results. With the multimodal information fusion module, the recommender system can more accurately capture user interests, increase the variety and interpretability of recommended content while meeting the demand for personalised recommendations, and provide a more efficient content distribution solution for short video platforms.

**2.3 Multimodal Information Fusion Module**

In order to ensure the security of user data privacy in short video recommendation systems, this paper designs a protection mechanism based on differential privacy [39]. The core of this mechanism is to scramble user data to ensure that the user's behavioral trajectory and sensitive information cannot be directly identified, while maintaining the efficiency and accuracy of the recommendation system. Figure 3 shows the system architecture of the differential privacy protection mechanism, which is divided into two parts: the mobile client and the server, responsible for data collection, protection, uploading, and processing.

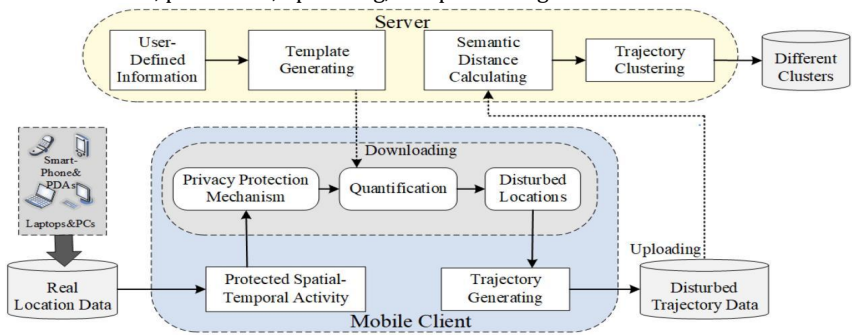

Figure 3: Differential Privacy Protection Mechanism Architecture in the Short Video Recommendation System

In mobile clients, user devices (e.g. smartphones, tablets or laptops) first collect real location data [40]. This location data is processed by privacy-preserving mechanisms, including perturbation and quantization, to generate protected spatio-temporal activity data. For example, the addition of noise corrupts the location data so that individual user data is no longer traceable. Perturbed trajectories are then generated using the protected data to ensure that the data uploaded to the server does not directly reflect the user's true activity trajectory. The server is primarily responsible for further processing and aggregating the uploaded data. First, the server generates templates based on user-defined information to standardise the trajectory data for different users. Then, by calculating the semantic distance, the system analyses the similarity of user activities, clusters the trajectory data, and generates clusters of trajectories of different categories [41]. This process effectively reveals the behaviour patterns of groups of users without revealing the specific behaviour of any individual user.

The main advantage of the differential privacy preserving mechanism is that it significantly improves privacy protection. By adding noise perturbations to the raw data, it prevents malicious attackers from reverse engineering real information about users. Despite the perturbations to the user data, the computation of semantic distance and trajectory clustering preserves the general trend of user behaviour, thus helping the recommender system to generate high-quality recommendations. In addition, the mechanism is highly scalable and can be adapted to different types of mobile devices and network environments, making it suitable for the privacy requirements of large-scale short video platforms. With the differential privacy preserving mechanism, the system ensures user privacy while maintaining the performance and user experience of the recommendation system. The mechanism provides important insights for the application of privacy-preserving techniques in future recommender systems.

## 3 ALGORITHM DESIGN AND OPTIMIZATION

### 3.1 Recommendation Algorithm Design

This paper proposes a recommendation algorithm that combines multimodal feature processing and differential privacy protection to enhance both the accuracy of recommendations and privacy protection. The recommendation algorithm consists of three parts: multimodal feature fusion, user interest matching, and the privacy protection mechanism, which will be elaborated below.Short video content contains multimodal information such as visual, textual, and audio features. To achieve efficient feature fusion, this paper introduces a weighted fusion model. Let the visual features $v_j$ of the video $v_j^{vis}$, the textual features be $v_j^{text}$, and the audio features be $v_j^{aud}$. The fused feature representation is given by Equation 1:

$$v_j = \alpha \cdot v_j^{vis} + \beta \cdot v_j^{text} + \gamma \cdot v_j^{aud} \quad (1)$$

where α, β, and γ represent the weights of each modality, which are dynamically adjusted through model training to optimally capture each modality's contribution to user interest. The fused feature vector $v_j$ expresses the multimodal features of the video and adapts to different user interests.

To quantify the user's interest in video $v_j$, this paper uses the matching score between the user feature vector $u_i$ and the video feature vector $v_j$ as the basis for recommendations. The matching score is calculated through a dot product operation and an activation function as shown in Equation 2:

$$s(u_i, v_j) = \sigma(u_i^\top v_j) \quad (2)$$

where σ(·) is the activation function, mapping the matching score to the range [0, 1], representing the user's probability of interest in the video. Based on the calculated matching scores, all videos are ranked from highest to lowest, and the top kk are selected as recommendations.To protect the privacy of user behavior data, the recommendation algorithm introduces differential privacy protection in the matching score calculation phase. Specifically, Laplace noise $s(u_i, v_j)$ is added to the matching score as shown in Equation 3:

$$s'(u_i, v_j) = s(u_i, v_j) + L(\frac{\Delta s}{\epsilon}) \quad (3)$$

where L(·) denotes the Laplace distribution, Δs s is the sensitivity of the matching score, and ϵ is the privacy budget parameter. By adding noise, the system can maintain the overall trend of recommendation results while protecting user privacy, balancing both privacy and practicality.With the above algorithm design, the recommendation system can fully utilize multimodal information to improve recommendation accuracy while safeguarding user data privacy, providing an efficient and reliable solution for short video recommendation platforms.

### 3.2   Differential Privacy Optimization Strategy

In short video recommendation systems, differential privacy protection mechanisms effectively prevent the leakage of user behavior data, but the addition of noise often impacts the performance of the recommendation algorithm. To balance privacy protection and recommendation effectiveness, this paper proposes a differential privacy optimization strategy based on dynamic noise adjustment. Traditional differential privacy mechanisms apply noise uniformly to all data, which may result in the excessive masking of key feature information, thereby reducing the accuracy of the recommendation system. This paper adopts a gradient-based importance-weighted method to dynamically adjust the noise strength based on the importance of data features. For higher-importance features, smaller noise is added to minimize the impact; for lower-importance features, noise is increased to enhance privacy protection.Let the matching score between the user $u_i$ and video $v_j$ be $s(u_i, v_j)$. To implement differential privacy protection, Laplace noise is added to the matching score. The optimization strategy dynamically adjusts the noise strength by introducing feature importance weights $\omega_{ij}$, and the optimized matching score formula is as shown in Equation 4:

$$s'(u_i, v_j) = s(u_i, v_j) + L(\frac{\Delta s}{\epsilon \cdot \omega_{ij}}) \quad (4)$$

Where $\omega_{ij}$ represents the feature importance weight, which is calculated based on gradient information as shown in Equation 5:

$$\omega_{ij} = \frac{|\nabla s(u_i, v_j)|}{max_k |\nabla s(u_i, v_k)|} \quad (5)$$

This reflects the gradient of the matching scores relative to the features, indicating the impact of these features on the final recommendation results. ϵ is the privacy budget and Δs s is the sensitivity of the matching scores. The main advantage of the dynamic noise adjustment strategy is that it focuses on protecting highly sensitive data, while maintaining the accuracy of the key feature recommendations. Experiments show that with the introduction of this optimisation strategy, the recommendation performance of the system is significantly improved compared to the traditional differential privacy mechanism, while maintaining the strength of user privacy protection. With this optimisation strategy, the recommender system can not only achieve efficient recommendations in complex multimodal computing environments, but also further reduce the negative impact of the privacy protection mechanism on performance, providing a safer and more reliable recommendation solution for short video platforms.

## 4 EXPERIMENTAL SETUP AND RESULTS ANALYSIS

### 4.1 Experimental Setup

The experiments were done on a server that had an Intel Xeon Gold 6226R 2.9GHz CPU, an NVIDIA Tesla V100 GPU, 128GB of RAM, and it was running Ubuntu 20.04. The deep learning framework used is PyTorch 2.0 and the privacy-preserving module is implemented using PySyft for differential privacy. The experiment uses public short video recommendation datasets, such as the YouTube-8M dataset, which includes information about the videos and the users' behaviour. The features we look at include:visual modality: features that tell us about the appearance of the video, such as what is in it and what is behind it. Textual modality: information from video titles, descriptions, and labels that reflect the meaning of the video. Audio modality: features extracted from audio spectrum analysis, such as speech, background noise and music. To evaluate the performance and privacy-preserving effect of the recommender system, we use the following key evaluation metrics: Precision@K: measures the proportion of relevant videos in the recommendation list. This metric shows how accurate the recommender system is and how well it can recommend relevant content based on the number of recommendations.Recall@K: Calculates the ratio of the number of clicks on relevant videos in the recommendation list to the total number of relevant videos, and evaluates how well the system covers the user's interests. Privacy Loss (Imperial Column): shows how private data is shared in the differential privacy protection mechanism, with lower values meaning better privacy protection. In our experiments, we adjust the privacy budget parameter ( $\varepsilon$ ) to control the strength of privacy protection and evaluate its impact on recommendation performance. Latency: shows the time it takes for the recommender system to respond to a request, and measures how quickly the recommender system responds in real time. To start the experiment, we take information from the dataset that includes images, words and sounds, and mix these together. We then use the test set to evaluate the models, calculating things like how accurate the recommendations are, how well they are recalled, how much privacy is lost, and how long it takes for the system to respond. The experiments also look at how the privacy budget ( $\varepsilon$ ) affects recommendation performance and explore the balance between privacy protection and how well recommendations work.

Table 1: Evaluation Metrics Table

| Metric Name | Description |
| --- | --- |
| Precision@K | Measures the proportion of relevant videos in the recommendation list, reflecting recommendation accuracy. |
| Recall@K | Measures the system's coverage of videos the user is interested in. |
| Privacy Loss ($\epsilon$) | Measures the strength of differential privacy protection, with smaller values indicating stronger privacy protection. |
| Latency | Measures the average response time of the recommendation algorithm, evaluating system real-time performance and usability. |

As shown in Table 1, this experiment provides a detailed evaluation of the performance of the short video recommendation system under multimodal feature fusion and privacy preservation mechanisms using the aforementioned settings and evaluation metrics. It also validates the performance of the proposed algorithm under different privacy budgets and provides valuable insights for privacy preservation and recommendation optimisation in real-world applications.

### 4.2 Experimental Results and Analysis

In this section, we present and analyze the experimental results of the short video recommendation system based on multimodal information and differential privacy protection. By comparing the recommendation performance and system effectiveness under different privacy budgets ($\epsilon$), we evaluate the impact of privacy protection strategies on the recommendation system. The results of the experiment show how well multimodal feature fusion and differential privacy optimisation strategies work in terms of recommendation accuracy, privacy preservation and system performance. The experiment was conducted with multiple rounds of evaluations under various privacy budgets, and the main experimental results are shown in Figure 4 below:

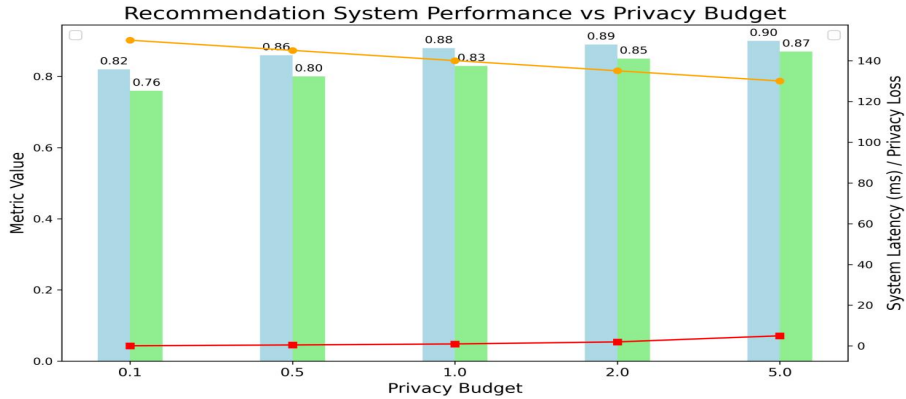

Figure 4: Experimental Results

As the privacy budget $\epsilon$ increases, both precision and recall gradually improve. This indicates that as the privacy protection strength decreases, the recommendation system becomes more accurate in recommending relevant videos to users. In cases with a higher privacy budget (e.g., $\epsilon=5.0$), the system can better utilize user behavior data for personalized recommendations, leading to improvements in both precision and recall. However, when the privacy budget is lower, the recommendation system is affected by noise, resulting in a decrease in recommendation performance.Privacy loss is positively correlated with the privacy budget. A lower $\epsilon$ value represents stronger privacy protection, but it also means the system must add more noise to protect user privacy. This results in a decrease in recommendation system performance under lower privacy budgets. In contrast, with a higher $\epsilon$¥epsilon, weaker privacy protection allows the system to access more raw data, improving recommendation performance.As the privacy budget increases, the system's response time slightly decreases. This is because the privacy protection mechanism requires the calculation and addition of noise. A higher privacy budget means that more data needs to be processed, which leads to longer computation time. However, the difference is minimal, and the system can still maintain low latency, ensuring real-time performance and usability of the recommendation system.

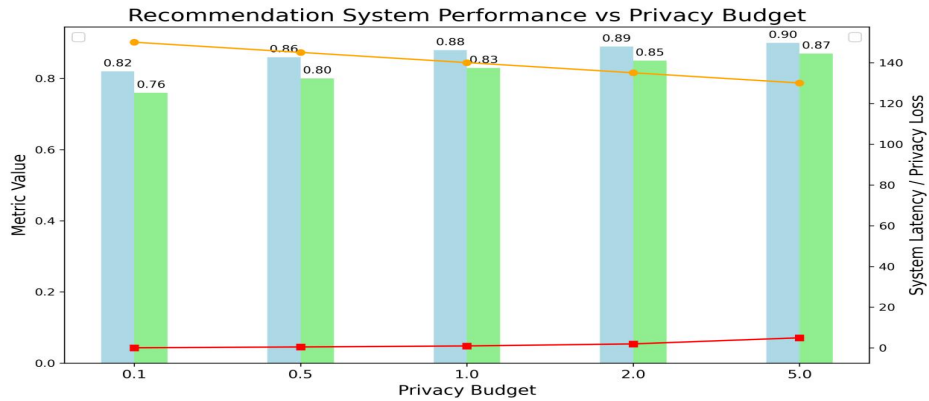

Figure 5: Experimental Results Analysis

Figure 5 shows that there is a trade-off between privacy budget and recommendation performance. A higher level of privacy may affect recommendation accuracy and recall, but it ensures that user privacy is not compromised. By dynamically adjusting the noise and optimisation strategies, the best balance can be found with different levels of privacy protection. The system's trade-off between privacy and recommendation accuracy provides flexibility for real-world privacy requirements and ensures the efficiency and reliability of the recommendation system.

## 5  CONCLUSION

This paper presents a short video recommendation system based on multimodal information and differential privacy protection. By combining multimodal feature fusion with differential privacy mechanisms, the system improves recommendation accuracy while ensuring user data privacy. Experimental results show that as the privacy budget ($\epsilon$) increases, the recommendation system's accuracy and recall improve, but the strength of privacy protection decreases, and the risk of user data leakage increases. Through optimization of the privacy protection strategy, an effective balance between privacy and recommendation performance can be achieved, providing a secure and efficient recommendation solution for short video platforms.

## Use of AI

Maybe somewhat ironically given that this is a perspective about the usefulness of AI, the intellectual work in this perspective is completely made by humans. However, ChatGPT 4 has been used as a writing assistance to suggest improvements of the grammar and the flow of the text, which is a great use for non-native English speakers.